\documentclass[conference]{IEEEtran}
\IEEEoverridecommandlockouts
\usepackage{cite}
\usepackage{amsmath,amssymb,amsfonts}
\usepackage{algorithmic}
\usepackage{graphicx}
\usepackage{textcomp}
\usepackage{xcolor}
\def\BibTeX{{\rm B\kern-.05em{\sc i\kern-.025em b}\kern-.08em
    T\kern-.1667em\lower.7ex\hbox{E}\kern-.125emX}}
\begin{document}

\title{Revisiting MFCCs: Evidence for Spectral-Prosodic Coupling \\
\thanks{This study was financed in part by the Coordenação de Aperfeiçoamento de Pessoal de Nível Superior – Brasil (CAPES) – Finance Code  001}
}

\author{\IEEEauthorblockN{Vitor Magno de O. S. Bezerra}
\IEEEauthorblockA{\textit{Electrical Engineering Department} \\
\textit{Federal University of Sergipe}\\
São Cristovão, Brazil \\
vitormagnosb@gmail.com}
\and
\IEEEauthorblockN{Gabriel F.A. Bastos}
\IEEEauthorblockA{\textit{Electrical Engineering Department} \\
\textit{Federal University of Sergipe}\\
São Cristovão, Brazil \\
gabrielfab210102@gmail.com}
\and
\IEEEauthorblockN{Jugurta Montalvão}
\IEEEauthorblockA{\textit{Electrical Engineering Department} \\
\textit{Federal University of Sergipe}\\
São Cristovão, Brazil \\
jmontalvao@academico.ufs.br}
}

\maketitle

\begin{abstract}
Mel-frequency cepstral coefficients (MFCCs) are an important feature in speech processing. A deeper understanding of their properties can contribute to the work that is being done with both classical and deep learning models. This study challenges the long-held assumption that MFCCs lack relevant temporal information by investigating their relationship with speech prosody. Using a null hypothesis significance testing framework, a systematic assessment is made about the statistical independence between MFCCs and the three prosodic features: energy, fundamental frequency (F0), and voicing. The results demonstrate that it is statistically implausible that the MFCCs are independent of any of these three prosodic features. This finding suggests that MFCCs inherently carry valuable prosodic information, which can inform the design of future models in speech analysis and recognition.
\end{abstract}


\section{Introduction} 
For decades, MFCCs have been a very important feature set in a wide range of speech processing applications \cite{nassif2019speech, wani2021comprehensive, hanifa2021review, o2024spoken, lopez2021deep}. MFCCs were designed to emulate the human auditory system by capturing spectral information according to the frequency-dependent critical bandwidths of the ear \cite{davis1980comparison}. Their ability to provide good acoustic discrimination and the low inter-correlation between coefficients made them particularly suitable for traditional statistical models \cite{anusuya2011front}. On the other hand, end-to-end speech recognition systems have achieved impressive results by learning features directly from raw waveforms or spectrograms \cite{mehrish2023review}. Therefore, the advent of these models has sparked a debate on the role of handcrafted features, such as MFCCs.


Nevertheless, MFCCs can also deliver state-of-the-art performance within deep learning-based frameworks, in architectures such as the Recurrent Neural Network Transducer (RNN-T) \cite{shi2021improving, lam2021raw}, which effectively integrates acoustic features from MFCCs with linguistic context modeling to perform speech recognition \cite{graves2012sequence}. Beyond state-of-the-art performance, MFCCs are also used in novel research across acoustic, industrial, and medical applications, often paired with classical machine learning models \cite{abdul2022mel}. The fact that MFCCs can support active research with both methodologies is relevant because it enables the framing of research questions that sidestep some challenges inherent to large-scale deep learning, such as ethical concerns \cite{reitmaier2022opportunities, slam2023frontier} and biological implausibility \cite{millet2022toward}. An example can be made with isolated word recognition, a task traditionally addressed using MFCCs with classical models that discard temporal dependencies between feature frames \cite{abdul2022mel, nassif2019speech}. However, this common practice raises the question: Could the temporal dynamics of the MFCCs also carry prosodic information?

MFCCs are spectral features traditionally used to model only the segmental effects of the vocal tract shape \cite{mehrish2023review, wani2021comprehensive}. This use is rooted in the linear source-filter theory, which posits that the periodic glottal source (airflow) is independent of the resonance of the vocal tract filter \cite{titze2008nonlinear}. According to this theory, the filter effects that are captured by spectral features are independent of the temporal variations of the source. These temporal variations are represented by another set of features that are related to speech prosody, features expressed mainly through three acoustic parameters: energy, fundamental frequency (F0), and duration \cite{mary2008extraction}. However, if the independence assumption is flawed, some of the prosodic information must be encoded inherently within spectral features such as MFCCs.

Combining information from the glottal source and the vocal tract filter is known to enhance performance across various speech applications \cite{wani2021comprehensive, teixeira2023narrative, hansen2015speaker}. MFCCs could inherently provide this complementary information if, contrary to common assumptions, they also encode prosodic features. While prior studies have recovered some prosodic information from MFCCs \cite{kim2020automatic, milner2006clean}, a systematic investigation of the extent of this relationship is lacking. This study addresses this gap by first establishing whether the assumption of source-filter independence can be statistically rejected. To accomplish this, we quantify the statistical dependence between MFCCs and prosodic features by estimating their conditional entropies. The significance of this dependence is then rigorously evaluated within a null hypothesis testing framework.

This paper is organized as follows. Section 2 details this procedure. Section 3 outlines the experimental setup, while Section 4 presents and discusses the results of the tests. Finally, Section 5 offers concluding remarks and outlines directions for future work. 

\section{Null Hypothesis Test} \label{sec: hypothesis_test}
As we mentioned above, the goal of this paper is to evaluate the assumption of independence between the MFCCs, which we denote by the random variable $X$, and some prosodic features, which we denote by the random variable $Y$, of an audio frame. This assumption is the null hypothesis $H_0$ that will be used to estimate how unlikely is this independence given a sequence of MFCCs  $S_{X}=\{x^{(i)}\}_{i=1}^{N}$ and a sequence of a prosodic feature $S_{Y_{\text{test}}}=\{y_{\text{test}}^{(i)}\}_{i=1}^{N}$ \cite{biau2010p}, where $x^{(i)}$ and $y^{(i)}$ encode, respectively, the MFCCs and the value of the prosodic feature from the $i$-th audio frame in a dataset. The degree of dependence between $X$ and $Y$ can be quantified in a variety of ways. In this work, we opted by using the information theoretical measure of conditional entropy $H(Y|X)$, which gives us the average amount of uncertainty about $Y$ that remains after knowing the variable $X$, as stated by Shannon \cite{shannon1948mathematical}. Therefore, when $X$ and $Y$ are dependent, $H(Y|X)$ is expected to be lowered compared to when $X$ and $Y$ are independent. This measure, when provided in bits (of information), informs that predicting the outcome of $Y$ given the outcome of $X$ is, on average, as difficult as predicting the outcome of an experiment with $C$ possible results uniformly distributed, where $C$ is effective cardinality \cite{montalvao2016minimum} of $Y$ given $X$, described by 
\begin{equation}
    C = 2^{H(Y|X)}.
\end{equation}
An assessment of the statistical significance of $C_\text{test}$ -- the effective cardinality estimated using sequences $S_{X}$ and $S_{Y_{\text{test}}}$ --, will show whether the evidence is against the null hypothesis or not. This assessment can be made with an empirical estimation of the distribution $p(C|H_0)$, which assumes that $H_{0}$ is true.

In order to estimate the conditional distribution $p(C|H_0)$, one must have access to an ensemble of effective cardinalities obtained from different sequences of samples from $X$ and $Y$ in which $H_0$ is known to be true. An easy way to obtain such sequences is to randomly shuffle the test sequence $S_{Y_{\text{test}}}$. Each permutation will produce a different sequence of prosodic features $S_{Y}$. Thus, each $x^{(i)}$ will be coupled with a random prosodic feature from another audio frame, ensuring the independence between $X$ and $Y$ in this estimation. Therefore, $p(C|H_0)$ can be estimated by calculating $C$ with $S_{X}$ and different permutations $S_{Y}$. The details on how to perform this procedure are explained in the remainder of the section.

The first step is to estimate the empirical probability of each possible discrete value of $X$. The sequence with $N$ elements is described by $S_{X} = \{x^{(1)}, x^{(2)}, ..., x^{(N)}\}$, where $x^{(i)} \in A$. Here, $A = \{a_1, a_2, ..., a_G\}$ is the set of discrete values that $X$ could assume.
The probability of $X$ assuming the value $a_j$ can be estimated as
\begin{equation}
    \hat{p}(X = a_j) = \frac{\sum_{i = 1}^{N}I_{x^{(i)} = a_j}}{N},
    \label{eq1}
\end{equation} 
where $I_{x^{(i)}
 = a_j}$ is defined by
\begin{equation}
    I_{x^{(i)} = a_j} =
    \begin{cases}
        1, & \text{if } x^{(i)} = a_j \\
        0, & \text{otherwise.}
    \end{cases}.
\end{equation}
 

The next step is to use a random permutation of the $N$ elements of $S_{Y_{\text{test}}}$ to create $S_{Y}$, a sequence given by $S_{Y} = \{y^{(1)}, y^{(2)}, ..., y^{(N)}\}$, where $y^{(i)} \in B$. Here $B = \{b_1, b_2, ..., b_L\}$ is the set of discrete values that $Y$ could assume. Analogously, the conditional probability $p(Y = b_l | X = a_j)$, also denoted by $p(b_l | X = a_j)$, of $Y$ assuming the value $b_l$ when $X$ has the value $a_j$ is estimated by 
\begin{equation}
    \hat{p}(b_l | X = a_j) = \frac{\sum_{i = 1}^{N}I_{y^{(i)} = b_l} \cdot I_{x^{(i)} = a_j}}{\sum_{i = 1}^{N}I_{x^{(i)} = a_j}}.
\end{equation}
The next step is to obtain estimates of the marginal conditional entropies $H(Y | X = a_j)$. Note that the sample sizes used to estimate these conditional entropies correspond to the amount of times each event of the form $X=a_j$ is observed in the dataset, which may be quite small for a proper statistical estimation. For this reason, we opted to use the Chao-Shen entropy estimator \cite{chao2003nonparametric}, which was consistently ranked among the most accurate estimators in experiments with varying sample sizes and domain sizes carried out in a comparative study in \cite{la2025bee}. The conditional entropies estimates using the Chao-Shen estimator are given by 
\begin{equation}
    \hat{H}(Y | X = a_j) = -\sum\limits_{l = 1}^{L}\frac{\hat{p}_{gt}(b_l | X = a_j)\log_{2}\hat{p}_{gt}(b_l | X = a_j)}{1-(1-\hat{p}_{gt}(b_{l}|X = a_{j}))^{n_{j}}},
\end{equation} where $n_j$ is the number of times $X=a_j$ is observed and $\hat{p}_{gt}(b_l | X = a_j)$ are the Good-Turing-corrected frequency estimates \cite{good1953population} given by 

\begin{equation}
    \hat{p}_{gt}(b_{l}|X = a_{j}) = \left ( 1-\frac{m_j}{n_j} \right ) \hat{p}(b_l|X = a_j),
\end{equation} where $m_j$ is the number of singletons in the sample, i.e. the number of events of the form $Y=b_l$ that were observed only once when $X=a_j$.

The average of all marginal conditional entropies of $Y$ is the conditional entropy $H(Y|X)$, which is estimated as
\begin{equation} \label{eq: conditional_entropy}
    \hat{H}(Y|X) = \sum\limits_{j = 1}^{M}\hat{p}(X = a_j)\hat{H}(Y|X = a_j).
\end{equation}
This measure can be used to estimate the effective cardinality $C$ using the equation 
\begin{equation}
    \hat{C} = 2^{\hat{H}(Y|X)}.
    \label{eq2}
\end{equation}
Thus, by repeating the above procedure for $D$ different permutations of $S_{Y_{\text{test}}}$, one is able to obtain a sequence of effective cardinalities $S_{C} = \{\hat{C}^{(1)}, ..., \hat{C}^{(d)}, ..., \hat{C}^{(D)}\}$ in which $H_{0}$ is known to be true. The values in this sequence are used in the empirical estimation of $p(C|H_0)$, given by
\begin{equation}
    \hat{p}(C|H_0) = \frac{\sum_{d = 1}^{D}I_{C^{(d)} = C}}{D}.
\end{equation}

By estimating the effective cardinality $C_{\text{test}}$ with the test sequence, we can also obtain an estimative of the probability of observing a result  as extreme as, or more extreme than, the one observed, under the null hypothesis, which is estimated as 
\begin{equation}
    \hat{p}(C \leq \hat{C}_{\text{test}} | H_{0}) = \frac{\sum_{d = 1}^{D}I_{C^{(d)} \leq C_{\text{test}}}}{D}.
    \label{eq3}
\end{equation}
This probability is called the $p$-value of the test \cite{fisher1958statistical}.

\section{Experimental Setup} 
This section details the methodology of the experiments, with information about the dataset, feature extraction, feature quantization, sequence organization, and testing procedure.

\subsection{Dataset}
The speech and laryngograph signals used in this study were sourced from a publicly available dataset \cite{bagshaw1994automatic}. This dataset comprises 50 English sentences spoken by both a male and a female speaker. Thus, the dataset contains a total of 100 audio signals sampled at a frequency $f_s = 20\mathrm{kHz}$. The selection of this corpus was motivated by three primary factors. Firstly, this dataset contains reliable F0 labels, thus dropping the need for using F0 estimators, which could bias our conclusions. Secondly, it has been previously utilized in the evaluation of F0 estimation algorithms, providing established benchmarks to be compared in future developments of this work \cite{xu2008pitch, camacho2008sawtooth}, in which we intend to develop F0 estimators using MFCCs information. Finally, the inclusion of both a male and a female speaker allows both genders to be considered in the tests, as the gender is known to influence the speech features -- it affects the performance of F0 estimators, for example.

\subsection{Feature Extraction}
The testing procedure described in the previous section requires a sequence of discrete values assotiated with the MFCCs $S_{X}=\{x^{(i)}\}_{i=1}^{N}$ and a sequence of discrete values $S_{Y_{\text{test}}}=\{y_{\text{test}}^{(i)}\}_{i=1}^{N}$ associated with each of the three prosodic features, namely the energy, the fundamental frequency (F0) and the voicing information. In this subsection, we describe how these four features are extracted, while the feature quantization procedures -- necessary to obtain sequences of discrete values -- are presented in the next subsection.


To extract the features, each speech signal in the dataset is divided into audio frames with a duration of $20\mathrm{ms}$ and $50\%$ overlap. That is, each frame is a discrete-time signal $s$ containing $M=400$ samples, and from each frame of the dataset, two raw features are extracted: the energy $e$, in logarithmic scale, and the set of 13 MFCCs $\{mfc_z\}_{z = 1}^{13}$. The energy can be calculated as \begin{equation}
    e^{(i)} = \ln{ \left (\sum\limits_{n = 1}^{M}s^{(i)}[n]^2 \right )},
\end{equation} where $s^{(i)}$ is the $i$-th audio frame. As for the MFCCs, they were calculated through the following five steps: 

\begin{itemize}
    \item [(i).] Pre-emphasis -- A first-order high-pass filter is applied to the signal to flatten the speech spectrum: $s_p[n] = s[n] - 0.97s[n-1]$.
    \item [(ii).] Windowing: A Hamming window is applied to the pre-emphasized frame to reduce spectral leakage: $s_w[n] = \left(0.54 - 0.46\cos\left(\frac{2\pi(n-1)}{M-1}\right)\right) \cdot s_p[n]$.
    \item [(iii).]  Discrete Fourier Transform (DFT): The windowed frame is zero-padded to 512 samples, and its magnitude spectrum is calculated as $F_k = \left| \sum_{n=0}^{511} s_w[n]e^{-j\frac{2\pi nk}{512}} \right|,$ where $k = 0, \dots, 511$.
    \item [(iv).]  Mel filterbank application: The magnitude spectrum is passed through a triangular filterbank composed of 23 filters, whose center frequencies $\{cf_h\}_{h=1}^{23}$, in Hz, are equidistant on the Mel scale. The output of the $h$-th filterbank is $fb_h = \sum_{l} W_h[l] F_l$, where $W_h[l]$ represents the triangular weighting of the $h$-th filter applied to the $l$-th frequency bin $F_l$.
    \item [(v).] MFCCs ($\{mfc^{(i)}_z\}_{z=1}^{13}$): The final coefficients are obtained by applying the Discrete Cosine Transform (DCT) to the logarithm of the filterbank energies: 
    \begin{equation}
         mfc^{(i)}_z = \sum_{h=1}^{23} \ln(fb_h) \cos\left(\frac{\pi z(h-0.5)}{23}\right),
    \end{equation}
    where $z = 1, \dots, 13$.
\end{itemize}

This procedure for MFCCs extraction is the same as the one used in \cite{abdul2022mel}.


Regarding the fundamental frequency and the voicing index, they were directly derived from the F0 labels provided within the dataset, which were originally extracted from the laryngograph signal at glottal closure instants. These labels are obtained at irregular time intervals, so the fundamental frequency of each frame $f0^{(i)}$ is obtained through a simple linear interpolation at every $10$ms, in alignment with each $e^{(i)}$ and each $\{mfc_z^{(i)}\}_{z=1}^{13}$, in every voiced region, and is assigned to zero in the unvoiced frames, yielding a frame-synchronous F0 contour. Finally, the voicing indexes are obtained as 
\begin{equation}
    v^{(i)} = 
    \begin{cases}
        1, & f0^{(i)} \neq 0\\
        0, & f0^{(i)} = 0
    \end{cases}.
\end{equation}

This is similar to the vowel/consonant indexing scheme used in \cite{kim2020automatic}, but instead of indexes pointing to vowel or consonantal frames, $v^{(i)}$ points to voiced or unvoiced frames. In this way, the voicing index acts as an \textit{ad hoc} measure of duration, an acoustic parameter associated with the rhythm of speech \cite{mary2008extraction}.

\subsection{Feature Quantization}
Out of the four features extracted, three of them have a continuous nature: the energy, the F0 and the MFCCs. Therefore, in order to model them as discrete random variables and use the approach proposed in section \ref{sec: hypothesis_test}, one must apply quantization to them. In this work, the energies $e^{(i)}$ and the F0's $f0^{(i)}$ were quantized through a simple rounding procedure, as this resolution was visually sufficient to keep both signals undistorted. As for the MFCCs $\{mfc_z^{(i)}\}_{z=1}^{13}$, a vector quantization method was necessary, since each MFCCs set is a 13-dimensional vector. The chosen approach was to train a Gaussian Mixture Model (GMM) with 40 components per speaker using the Expectation-Maximization (EM) algorithm \cite{bishop}, as it was experimentally verified that further increasing the number of components above 40 does not cause significant improvements in the model's likelihood. Then, each MFCCs vector $\{mfc^{(i)}_z\}_{z=1}^{13}$ was replaced by a single integer index $id^{(i)}$ corresponding to the Gaussian component having the highest posterior probability for that vector.

Since GMM-based quantization of MFCCs is known to encode speaker identity \cite{reynolds2000speaker}, combining data from multiple speakers would create a prior dependency between feature sequences, since a speaker identity can also be associated with prosodic information \cite{mary2008extraction}. To prevent this, the quantization and the analysis processes were conducted for each speaker independently.

\subsection{Sequence Preparation}
For each speaker, every audio signal is associated with four sequences of quantized features -- one set of feature values for each frame --. The final sequences of each speaker are obtained by concatenating the sequences associated with every one of the 50 audio signals. In each test run, three types of sequences were utilized: a sequence of MFCCs indexes in its natural chronological order ($S_X$), a corresponding prosodic feature sequence also in its natural order ($S_{Y_{\text{test}}}$), and a collection of $D$ randomly shuffled versions of the prosodic sequence ($S_Y$). When either the energy or the F0 was used as the prosodic feature, all unvoiced frames -- the ones with $f0^{(i)} = 0$ -- were removed from all sequences to ensure that the analysis was restricted to voiced regions. This setup allows for a direct evaluation of dependencies in naturally ordered sequences $S_{Y_{\text{test}}}$ and $S_X$ using the proposed approach, with a distribution $\hat{p}(C|H_0)$ obtained from the randomly ordered sequences.

\subsection{Testing Procedure}
In total, six tests were performed -- one for each combination of prosodic feature and speaker. All tests were performed following the four steps outlined below: 
\begin{itemize}
    \item [(i).] Selection of which effective cardinality $\hat{C}^{(feat.,gen.)}_\text{test}$ will be evaluated in the test, where $gen.$ (gender) is either $fem.$ (female) or $mal.$ (male) and $feat.$ (feature) is either $e.$ (energy), $f0$ (F0) or $v.$ (voicing).
    \item [(ii).] From the sequence $S_{X} = \{x^{(i)}\}_{i=1}^{N} = \{id^{(i)}\}_{i=1}^{N}$ containing the quantized MFCCs of the selected gender, the probabilities $\hat{p}(X = a_j)$ are estimated using \eqref{eq1}. 
    \item [(iii).] From the sequence $S_{Y_{\text{test}}} = \{y_\text{test}^{(i)}\}_{i=1}^{N}$ containing the quantized values of the chosen prosodic feature from the selected gender, the effective cardinalities under the null hypothesis are obtained from $D= 10^5$ different permutations of $S_{Y_{\text{test}}}$ using \eqref{eq2}. In this step, the effective cardinality $\hat{C}_{\text{test}}$ is also obtained from the natural sequence $S_{Y_\text{test}}$.
    \item [(iv).] Finally, an upper bound for the $p$-value is obtained using \eqref{eq3}.
    
\end{itemize}

\section{Results and Discussion}
\begin{figure}[tbp!]
\centerline{\includegraphics[width=0.33\textwidth]{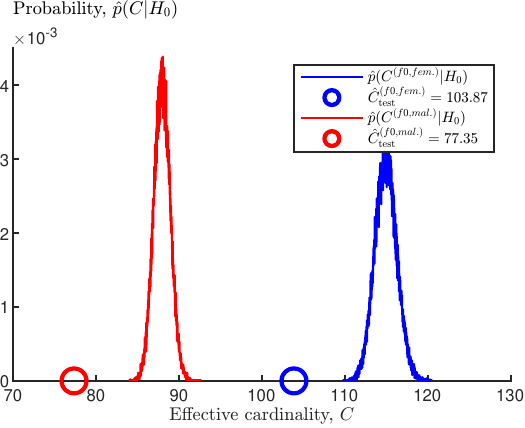}}
\caption{Empirical distributions, $\hat{p}(C|H_0)$, of the effective cardinality under the null hypothesis ($H_0$). The distributions obtained from $S_Y = \{f0^{(i)}\}_{i = 1}^{N}$ are shown for female (blue) and male (red) speakers. The test effective cardinalities of 103.87 (female) and 77.35 (male) are indicated by circle markers.}
\label{fig1}
\end{figure}

\begin{figure}[tbp!]
\centerline{\includegraphics[width=0.33\textwidth]{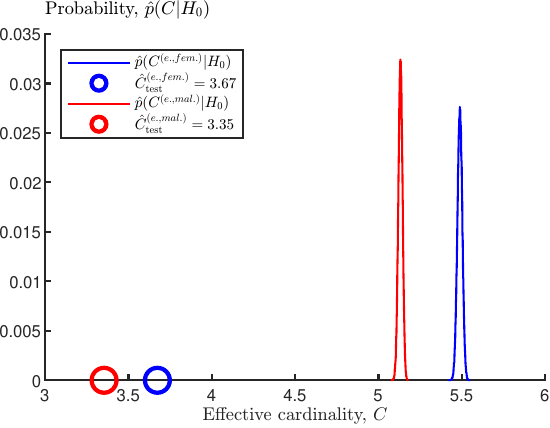}}
\caption{Empirical distributions, $\hat{p}(C|H_0)$, of the effective cardinality under the null hypothesis ($H_0$). The distributions obtained from $S_Y = \{e^{(i)}\}_{i = 1}^{N}$ are shown for female (blue) and male (red) speakers. The test effective cardinalities of 3.67 (female) and 3.35 (male) are indicated by circle markers.}
\label{fig2}
\end{figure}

\begin{figure}[hbtp!]
\centerline{\includegraphics[width=0.33\textwidth]{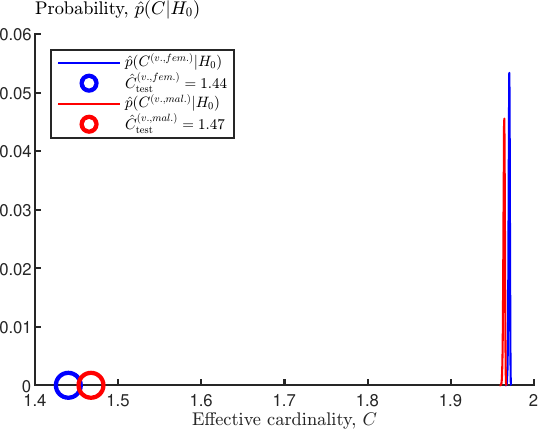}}
\caption{Empirical distributions, $\hat{p}(C|H_0)$, of the effective cardinality under the null hypothesis ($H_0$). The distributions obtained from $S_Y = \{v^{(i)}\}_{i = 1}^{N}$ are shown for female (blue) and male (red) speakers. The test effective cardinalities of 1.44 (female) and 1.47 (male) are indicated by circle markers.}
\label{fig3}
\end{figure}

In all six tests, the null hypothesis ($H_0$) was rejected with high level of significance. This occurred because, in every test, a result as extreme as, or more extreme than the effective  cardinality observed in the naturally ordered sequence, $\hat{C}^{(feat.,gen.)}_\text{test}$, was never produced under the null hypothesis. This suggests $p$-values below $10^{-5}$, as an occurence of these events were to be expected in $10^5$ trials if their probability was higher than $10^{-5}$. These results, shown for both genders across the F0, energy, and voicing features in Fig.~\ref{fig1}, Fig.~\ref{fig2} and Fig.~\ref{fig3}, respectively, strongly suggest that the observed reduction in effective cardinality is due to prosodic information contained within the MFCCs. However, the fact that the effective cardinality remains large indicates  that while the MFCCs might be relevant for prosodic analysis, they alone may not be enough to fully access this information.


\section{Conclusion}
Despite their long-standing use in audio processing, the full potential of MFCCs remains unrealized. This work challenged the long-held assumption of their independence from prosodic information by introducing a novel null hypothesis testing procedure. Our results conclusively reject this assumption, providing clear evidence that MFCCs do, in fact, contain significant prosodic information. Therefore, we believe that the development of methods to effectively extract this information using MFCCs is a promising avenue for future research. More specifically, we intend to investigate whether the remaining uncertainty about the prosody after knowing the MFCCs, found in the results of the previous section, can be removed with the use of contextual information, i.e. using a window of MFCCs from the neighboring frames, as opposed to only the MFCCs from the target frame. In a sense, this study underscores the continued relevance of MFCCs, highlighting the need for further research to unlock their untapped capabilities.

\bibliographystyle{IEEEtran}
\bibliography{bibliografia}

\end{document}